\documentclass[aps,prl,twocolumn]{revtex4}
\usepackage{amssymb}
\usepackage{amsmath}
\usepackage{epsfig}
\begin{document}

\title{All optical Amplification in Metallic Subwavelength Linear Waveguides}

\author {Ramaz Khomeriki${}^{1,2}$,  J\'er\^ome Leon${}^3$\footnote{Prof. J\'er\^ome Leon passed away before publishing the paper}}
\affiliation{${\ }^1$Max-Planck Institute for the Physics of
Complex Systems, N\"othnitzer Str. 38, 01187 Dresden (Germany) \\
${\ }^2$Physics Department, Tbilisi State University, 0128
Tbilisi (Georgia) \\ ${\ }^3$Laboratoire de Physique Th\'eorique et Astroparticules \\
CNRS-IN2P3-UMR5207, Universit\'e Montpellier 2, 34095 Montpellier
(France)}


\begin{abstract}
Proposed all optical amplification scenario is based on the
properties of light propagation in two coupled subwavelength
metallic slab waveguides where for particular choice of waveguide
parameters two propagating (symmetric) and non-propagating
(antisymmetric) eigenmodes coexist. For such a setup incident
beams realize boundary conditions for forming a stationary state
as a superposition of mentioned eigenmodes. It is shown both
analytically and numerically that amplification rate in this
completely linear mechanism diverges for small signal values.
\end{abstract}

\maketitle

Leading ideas in investigations of all optical logical devices in
structured media \cite{christo} usually implement optical
bistability \cite{bi,ramaz} or soliton interaction \cite{mcleod}
in creating switching operation of optical beams. Quantum dots
\cite{dot}, single molecule \cite{mol} or atomic systems
\cite{atom} could be also used for optically controlled switching
of light. One can also quote various optoelectronic approaches
\cite{opto} for realization of optical transistors and asymmetric
nonlinear waveguides for all optical diodes \cite{lepri}. However,
all the mentioned setups are based on nonlinear photon-photon
interactions and hardly meet the criteria \cite{photo} for
applicability in all optical computing. Here we consider the
possibility of amplification of optical signals using two
subwavelength waveguides coupled by metallic film. The problem is
linear with no need in high power fields and there is a reach
experience in building of subwavelength photonic \cite{segev} and
metallic waveguides \cite{metal}.

%%%%%%%%%%%%%%%%%%%%%%%%%%%%%%%%%%%
\begin{figure}[t] \centerline
{\epsfig{file=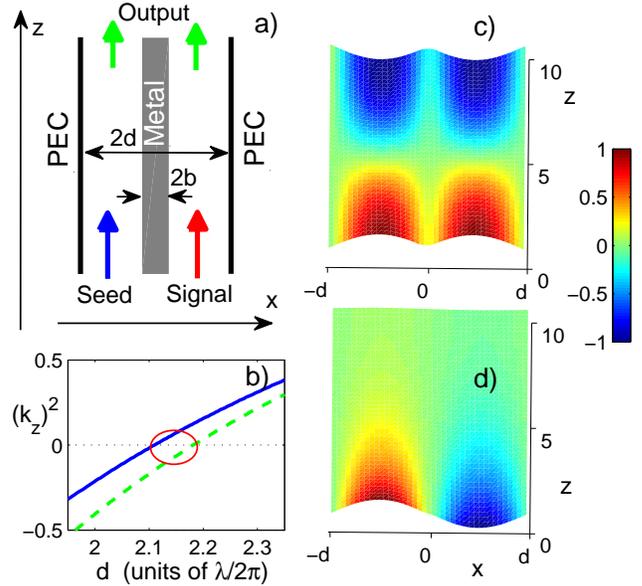,width=8.5cm}} \caption{(Color online). a)
Schematics for dielectric-metal-dielectric waveguide system
restricted by perfect electric conductor (PEC) from both sides. b)
dependence of longitudinal wavenumber on the waveguide width $d$
for two fundamental symmetric (solid) and antisymmetric (dashed)
modes calculated form Eqs. \eqref{disp} and \eqref{cont};
thickness of metallic film is fixed to the value
$2b=0.506\cdot(\lambda/2\pi)$. c) and d) display snapshots for
these modes according to Exps. \eqref{solpm} and \eqref{cond11}.}
\label{fig:map}\end{figure}
%%%%%%%%%%%%%%%%%%%%%%%%%%%%%%%%%%%%%%%%

Our idea of all optical amplification is based on the possibility
of coexistence of two fundamental modes with real (symmetric mode)
and imaginary (antisymmetric mode) wavenumbers along longitudinal
propagation direction of dielectric-metal-dielectric combined
waveguide system. In such a situation only symmetric mode can
carry nonzero flux, while the energy flux associated with
antisymmetric mode is exactly zero. Thus the propagation of
antisymmetric mode responsible for destructive interference is
suppressed and amplification effect can take place.

Suggested effect is different than one in a homodyne receiver
scheme, where signal intensity is amplified at the receiver area,
while the total signal energy flux is not amplified. As we will
show below, the total signal flux amplification is possible only
in metallic subwavelength waveguides, where symmetric and
antisymmetric modes are characterized by real and imaginary
wavenumbers, respectively.

In principle, proposed waveguide system could be of different
geometries, in this paper we just consider two dielectric slabs
(with refractive index $n$) separated by metallic film and we
assume perfect electric conductor (PEC) condition at both sides of
waveguide system. Thus the setup presented in Fig. \ref{fig:map}a
just allows to reduce the problem to 2 (space) + 1 (time)
dimensions assuming the electric field polarized and homogeneous
along $y$ direction and having fixed zero value at the boundaries.
In nonmagnetic medium we can write down the following wave
equations for the transverse electric field $E\equiv E_y$
perpendicular to $xz$ plane:
\begin{equation}\label{Max-base}
\Delta{E}-n^2\partial_{tt}{E}=0, \quad
\Delta{E}-\omega_p^2E-\partial_{tt}{E}=0,
\end{equation}
where we work in the units $c=1$ and a definition
$\Delta\equiv\partial_{xx}+\partial_{zz}$ is introduced. First
equation in \eqref{Max-base} corresponds to the wave propagation
in a dielectric, while the second describes dynamics inside
metallic film in approximation of zero Drude relaxation rate.
Beyond this approximation electromagnetic wave dynamics inside
metal is governed by \cite{ash}
\begin{equation}
\Delta{E}-4\pi\partial_tJ-\partial_{tt}{E}=0, \qquad \partial_t
J=-\frac{J}{\tau}+\frac{\omega_p^2}{4\pi}E \label{dop}
\end{equation}
where $J$ stands for the electric current density, $1/\tau$ stands
for Drude relaxation rate, $\omega_p=\sqrt{4\pi e^2N/m}$ is a
plasma frequency; $e$, $N$ and $m$ are charge, concentration and
mass of electrons, respectively.

For the sake of analytical simplifications we assume negligible
damping inside the metal ($\tau\rightarrow \infty$) getting from
\eqref{dop} automatically the initial system \eqref{Max-base}, and
we will work in nontransparent for the metal frequency range
$\omega\ll\omega_p$. Moreover, because of the placement of PEC at
both sides of the waveguide system, one has vanishing boundary
conditions $E(x=\pm d)=0$. Thus stationary basic solution of
\eqref{Max-base} in the different parts of the combined
dielectric-metal-dielectric symmetric waveguide system is written
as follows:
\begin{align}
&E=A\sin\left[k_x (d+x)\right]e^{i(k_z z-\omega t)}+c.c. \qquad
-d<x<-b \nonumber
\\ &E=\left(F_1 e^{\kappa x}+F_2e^{-\kappa x}\right)e^{i(k_z z-\omega t)}
+c.c. \qquad |x|<b \label{cond1}\\
&E=B\sin\left[k_x (d-x)\right]e^{i(k_z z-\omega t)}+c.c. \qquad
b<x<d, \nonumber
\end{align}
where "c.c." means complex conjugated term. Here we assume that in
the combined part of the waveguide one has a dielectric in the
range $b<|x|<d$ and metal in the range $|x|<b$; $A$, $B$, $F_1$
and $F_2$ are real amplitudes of electric field in dielectric and
metallic parts, respectively, $k_x$ is a real wavenumber in the
dielectric, $\omega$ is a working frequency and $\kappa$ is
penetration depth in the metal. These wavenumbers are linked by
the dispersion relations
\begin{eqnarray}
k_z=\sqrt{\omega^2n^2-(k_x)^2}, \quad
\kappa=\sqrt{\omega_p^2-\omega^2+(k_z)^2}, \label{disp}
\end{eqnarray}
which automatically follows putting solution \eqref{cond1} into
wave equations \eqref{Max-base}. Fixing operational frequency
$\omega$ and waveguide parameters $b$ and $d$, all other
quantities are uniquely defined. Particularly, from the continuity
conditions of solution \eqref{cond1} at the lines $x=\pm b$ one
gets following relations for $k_x$:
\begin{equation} \label{cont}
\tan\left[k_x^{\pm}(d-b)\right]\left[\tanh(\kappa b)\right]^{\pm
1}+\left[k_x^{\pm}/\kappa\right]=0,
\end{equation}
where we have $+$ ($-$) sign for symmetric (antisymmetric)
solution. Taking into account dispersion relations \eqref{disp}
one can calculate $k_x^\pm$ and $k_z^\pm$ versus waveguide
parameters $b$ and $d$ and we are interested in the range of these
parameters for which $\left(k_z^+\right)^2$ is positive while
$\left(k_z^-\right)^2$ is negative (see Fig. \ref{fig:map}b for
the appropriate parameter values indicated by a circle). Then
defining real quantities as $k_z^+\equiv k_s$ and $k_z^-\equiv
ik_a$ we can write for symmetric and antisymmetric solutions:
\begin{equation}
E_s=\Phi_+(x)\cos(k_sz-\omega t); \qquad E_a=\Phi_-(x)e^{-k_a
z}\cos(\omega t), \label{solpm}
\end{equation}
where orthogonal to each other symmetric and antisymmetric
profiles $\Phi_{\pm}(x)$ are defined as:
\begin{align}
&\Phi_{\pm}=\sin\left[k_x^\pm (d+x)\right] \qquad -d<x<-b
\nonumber
\\ &\Phi_{\pm}=-\frac{\sin[k_x^\pm(d-b)]}{e^{\kappa b}\pm e^{-\kappa b}}
\left[e^{\kappa x}\pm e^{-\kappa x}\right]
 \qquad |x|<b \label{cond11}\\
&\Phi_{\pm}=\pm\sin\left[k_x^\pm (d-x)\right] \qquad b<x<d
\nonumber
\end{align}
%%%%%%%%%%%%%%%%%%%%%%%%%%%%%%%%%%%
\begin{figure}[t] \centerline
{\epsfig{file=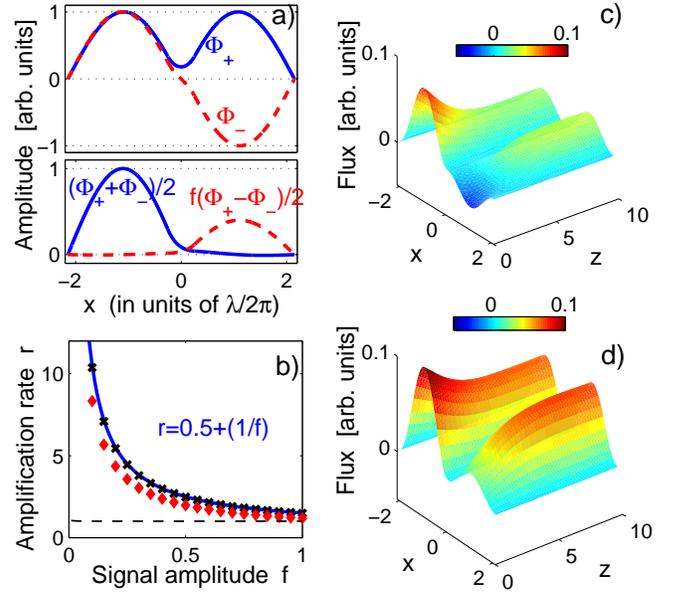,width=8.5cm}} \caption{(Color online). a)
upper panel displays profiles of two fundamental modes given by
Exps. \eqref{cond11} and lower panel presents their combinations
which serve as a good approximation for boundary conditions of
beams entering into left (solid line) and right (dashed line)
waveguides. b) results of numerical simulations on amplification
rate versus signal amplitude $f$ in cases of zero relaxation rate
(crosses) and realistic relaxation rate for silver (diamonds),
which is compared with analytical formula \eqref{rate} presented
as a solid line. Dashed line displays amplification rate for
non-subwavelength waveguide (waveguide width is doubled) following
the formula \eqref{rate1}. c) and d) show energy flux density
distribution within the waveguide system for signal amplitudes
$f=0$ and $f=0.5$, respectively. In numerical simulations the
following waveguide parameters are utilized: waveguide total width
$2d=4.294$ and metallic film thickness $2b=0.506$, these
parameters together with spatial coordinates $x$ and $z$ are in
units of $\lambda/2\pi$, where $\lambda=0.7\mu m$ is a vacuum
wavelength of operating laser frequency.} \label{fig2}\end{figure}
%%%%%%%%%%%%%%%%%%%%%%%%%%%%%%%%%%%%%%%%
and we present snapshots of symmetric and antisymmetric solutions
\eqref{solpm} in Fig. \ref{fig:map}c,d while their profiles
\eqref{cond11} along axis $x$ and their combinations are presented
in Fig. \ref{fig2}a.

As far as electric field has a single component along transversal
$y$ axis one can readily compute in-plane components of magnetic
field, particularly, $H_x$ could be easily integrated form the
Maxwell equation $\partial H_x/\partial t=\partial E/\partial z$.
Then it is straightforward to calculate energy flux density as
$s_z=EH_x$ and total energy flux along longitudinal $z$ direction
$S_z=\int_{-d}^d EH_xdx$. It is easy to see from \eqref{solpm} and
\eqref{cond11} that averaged over time total flux in case of
symmetric eigenfunction is $\langle S_z^s\rangle \simeq
dk_s/2\omega$, while in antisymmetric case one has a standing wave
profile along $z$ direction and consequently averaged total flux
$\langle S_z^a\rangle$ is exactly zero.

Now the question is which solution (symmetric or antisymmetric one
or their linear combination) is realized for the given boundary
condition. Let us suppose that seed and input beams are injected
from the isolated waveguides separated by perfect electric
conductor (PEC). Thus we have the seed and input waveguides
bounded by PEC at $x=-d,0$ and $x=0,d$, respectively, and first of
all we consider symmetric incident field in the form of the
following propagating wave at $z<0$:
\begin{equation} \label{ss}
I_s=\left|\sin\left(\pi x/d\right)\right|\cos(k_sz-\omega t)
\qquad -d<x<d.
\end{equation}
This should be combined with the reflected beam with unknown
amplitudes $r_1$ and $r_2$ characterizing symmetric and
antisymmetric contributions
\begin{equation} \label{ss1}
R=\left[r_1\sin(\pi x/d)+r_2\left|\sin\left(\pi
x/d\right)\right|\right]\cos(k_sz+\omega t)
\end{equation}
and the sum $I_s+R$ should be connected with the linear
combination $u_sE_s+u_aE_a$ of the solutions at $z>0$ given by
Exp. \eqref{solpm} via continuity conditions. Noting that in case
of narrow metallic films $b\ll d$ (see Fig. \ref{fig2}a)
$\Phi_+\simeq |\sin(\pi x/d)|$ and $\Phi_-\simeq -\sin(\pi x/d)$
we easily get the condition $r_1=r_2=u_a=0$, $u_s=1$ meaning that
there is no reflected wave and in a whole range of $z$ and the
symmetric solution is approximately given by:
\begin{equation} \label{ss0}
E_s\simeq\left|\sin\left(\pi x/d\right)\right|\cos(k_sz-\omega t).
\end{equation}

While in case of antisymmetric incident field the analysis is a
bit more complicated, particularly, if we take the incident field
in the form
\begin{equation} \label{aa}
I_a=-\sin\left(\pi x/d\right)\cos(k_sz-\omega t) \qquad -d<x<d
\end{equation}
from the similar to above continuity considerations we conclude
that such a beam is completely reflected and a whole solution is
written as follows:
\begin{align}
&E_a\simeq -\sin(\pi x/d)\left[\cos(k_sz-\omega
t)+\cos(k_sz+\omega
t+\varphi)\right] \nonumber \\
&E_a\simeq -u_a\sin(\pi x/d)e^{-k_az}\cos(\omega t+\varphi/2)
\label{aa1}
\end{align}
for $z<0$ and $z>0$, respectively, and the costants are defined as
$\tan(\varphi/2)=k_a/k_s$ and $u_a=2/\sqrt{1+(k_a/k_s)^2}$.

Now let us suppose that we have an incident seed beam entering
into the left waveguide and this could be presented as a linear
combination of symmetric \eqref{ss} and antisymmetric \eqref{aa}
incidences:
\begin{equation} \label{seed}
I=\left[|\sin(\pi x/d)|-\sin(\pi x/d)\right]\cos(k_sz-\omega
t)=\left(I_s+I_a\right)/2
\end{equation}
and as a consequence a whole solution under such a boundary
conditions is given as linear combination of complete solutions
\eqref{ss0} and \eqref{aa1} (see also bottom panel of Fig.
\ref{fig2}a):
\begin{equation}\label{prim}
E\simeq\left(E_s+E_a\right)/2.
\end{equation}

From \eqref{seed} it is easy to calculate total averaged in time
flux carried by seed beam entering into the left waveguide:
$\langle S_z^I\rangle=k_sd/4\omega$. At the same time, calculating
the energy flux of the complete solution and taking into account
that the flux of antisymmetric solution is strictly zero, one gets
half of the value of incident averaged flux: $\langle S_z^I\rangle
/2=k_sd/8\omega$, meaning that only half of the incident intensity
goes through the waveguide system and the rest is reflected back.
As we will show below, by injecting signal beam (having the same
phase as the seed beam) into the right waveguide reflected flux
gradually decreases increasing amplitude of the signal beam and
reaches zero value when both input and seed beam have equal
amplitudes. This is a reason for amplification mechanism of the
signal beam. For quantitative description of this effect we
perform the similar analysis for the incident signal field with
amplitude $f$, particularly, presenting it as
\begin{equation} \label{sig}
I^f=f\left[|\sin(\pi x/d)|+\sin(\pi x/d)\right]\cos(k_sz-\omega
t).
\end{equation}
This incident field carries the averaged total flux
\begin{equation} \label{flsig}
\langle S_z^f\rangle=f^2k_sd/4\omega
\end{equation}
which is a total gain of incident flux due to application of the
signal field. Thus in case of application of both seed
\eqref{seed} and signal \eqref{sig} beams the total incidence is
$I+I^f$ and it realizes the complete solution in the form:
\begin{equation}\label{solf}
E\simeq\frac{(1+f)E_s}{2}+\frac{(1-f)E_a}{2}.
\end{equation}
From the similar to above arguments that antisymmetric mode is
characterized by a zero averaged flux, it is obvious that such a
solution carries the averaged flux $\langle
S_z^{out}(f)\rangle=(1+f)^2k_sd/8\omega$ and thus total gain of
the output flux due to the signal is read as
\begin{equation}
\langle S_z^{out}(f)\rangle-\langle
S_z^{out}(0)\rangle=\frac{(2f+f^2)k_sd}{8\omega}
\end{equation}
and this should be compared with averaged over time input signal
flux \eqref{flsig}, and thus one gets for the amplification rate:
\begin{equation}
r=\frac{\langle S_z^{out}(f)\rangle-\langle
S_z^{out}(0)\rangle}{\langle S_z^f\rangle}=\frac{1}{2}+\frac{1}{f}
\label{rate}
\end{equation}
which diverges at small signal values.

It should be especially emphasized that such amplification of
signal beam takes place only in case of subwavelength waveguides
when the wavenumber of antisymmetric solution takes imaginary
value. In case of non-subwavelenth waveguides both symmetric and
antisymmetric solutions are characterized by real wavenumbers
which have very close values. Defining real wavenumber of
antisymmetric mode as $k^\prime_a\equiv k_z^-$ we note that
antisymmetric mode is also characterized by nonzero flux, and from
\eqref{solf} now we get a following expression for the
amplification rate:
\begin{equation}
r=\frac{k_s+k^\prime_a}{2k_s}+\frac{1}{f}\frac{k_s-k^\prime_a}{k_s}.
\label{rate1}
\end{equation}
For instance, if one has twice wider waveguide system the
coefficient at the divergent term is negligible
$(k_s-k^\prime_a)/2k_s=0.0005$ making amplification mechanism
ineffective: $r\simeq 1$ (see dashed line in Fig. \ref{fig2}b).

Next our aim is to confirm this analytical result by numerical
simulations. For this purpose we should first derive the
boundary-value data at the lines $z=0$ and $z=L$ ($L$ is a length
of the system) following Ref. \cite{chen,jerome}, that represent
incident waves entering the combined waveguide from the seed and
signal and going out. As we have mentioned above before entering
the waveguide system the seed and signal fields are described by
Exps. \eqref{seed} and \eqref{sig}, thus in the range $z\le 0$ the
solution reads
\begin{equation}\label{z0}
E=\left[I(x)\cos(k_sz-\omega t)+R(x)\cos(k_s z+\omega t)\right]
\end{equation}
where one has for $I(x)$
\begin{equation}\label{II}
I(x)=-\sin\left[\pi x/d\right] \quad \mbox{and} \quad
I(x)=f\sin\left[\pi x/d\right]
\end{equation}
for $-d<x<0$ and $0<x<d$, respectively, while $R(x)$ is unknown
amplitude profile for reflected wave and has thus to be
eliminated. The continuity conditions at $z=0$ with the electric
field $E(x,z,t)$ inside the combined waveguide can be written as
\begin{align*}
\left[I(x)+R(x)\right]\cos(\omega t)= \big(E\big)_{z=0}, \\
k_s\left[I(x)-R(x)\right]\sin(\omega
t)=\big(\partial_zE\big)_{z=0},
\end{align*}
which can be combined to exclude the unknown reflected amplitude
$R(x)$ taking time derivative from the first equation and then
combine resulting one with the second equation. The similar
manipulations could be done with boundary conditions at $z=L$ but
now there is no contribution of backward propagating field, thus
the resulting equations read as
\begin{align}
&\partial_zE\Big|_{z=0}=(k_s/\omega)\partial_tE\Big|_{z=0}+2k_s
I(x)\sin(\omega t),\nonumber \\
&\partial_zE\Big|_{z=L}=-(k_s/\omega)\partial_tE\Big|_{z=L}.
\label{bound1}
\end{align}
Thus we solve numerically initial equations \eqref{Max-base} with
boundary conditions \eqref{bound1} and definition \eqref{II} for
$I(x)$. Next we compute averaged over time longitudinal flux
density $\langle s_z\rangle$ inside the waveguide system and its
total value $\langle S_z\rangle$ across the system and compare the
latter to the value of the total incident signal flux given by
Exp. \eqref{flsig} for different values of signal amplitude $f$.
Finally we compare numerical results with the analytical
prediction \eqref{rate}. We choose operational frequency $\omega$
such that vacuum wavelength is $\lambda=0.7\mu m$, as a dielectric
we take glass with refractive index $n=1.5$ and we choose Silver
as a metal. Its complex refractive index for the mentioned
wavelength is \cite{silver}
\begin{equation}
\tilde n=n_1+in_2=0.05+5i, \label{sil}
\end{equation}
thus we can derive plasma frequency needed in \eqref{Max-base} as
follows \cite{born} $\omega_p\simeq n_2\omega$. The width of the
waveguide system is taken as $2d=4.294\cdot(\lambda/2\pi)$ and
metallic film thickness is chosen as
$2b=0.506\cdot(\lambda/2\pi)$. For such a choice of waveguide
parameters the wavenumber of symmetric propagating mode is
$k_s=0.25\cdot(2\pi/\lambda)$ and this value is used in boundary
conditions \eqref{bound1}. First of all we proceed with a
simplified case of zero relaxation rate $1/\tau=0$: Measuring
total flux for various values of signal amplitudes we have plotted
Fig. \ref{fig2}b which shows an excellent correspondence with
analytical formula \eqref{rate}, while in Fig. \ref{fig2}c,d we
plot the distribution of averaged in time flux density for two
values of signal amplitude $f=0$ and $f=0.5$. Finally, in
supporting material the time evolution animations of electric
field and associated flux densities are presented.

Next we made numerics using Eqs. \eqref{dop} and calculating Drude
relaxation rate from \eqref{sil} applying the formula
$1/\tau\simeq 2n_1\omega/n_2$. We use the same waveguide
parameters as in case of zero relaxation rate and display the
results for amplification rate in Fig. \ref{fig2}b (diamonds), as
seen even in this case the results are in good agreemnetg with
analytical formula \eqref{rate}.

Concluding, we present novel mechanism of signal amplification
based solely on linear effects and confirm the amplification
scenario by numerical simulations. In principle the analysis could
be extended in case of single waveguide with metallic boundary
when the seed is directly injected into the waveguide, while the
signal beam is illuminated from the metallic film side. The above
studies could be generalized for different systems where
propagating and nonpropagating fundamental modes coexist.

\acknowledgements{R. Kh. is indebted to S. Flach, F. Lederer, T.
Pertsch and A. Szameit  for many criticisms and useful
suggestions. The work is supported by joint grant from CNRS and
SRNSF (grant No 09/08). and SRNSF grant No 30/12.}

\end{document}